\documentclass[letterpaper, 10pt, journal, twoside]{IEEEtran}
\usepackage[utf8]{inputenc}

\usepackage{graphicx}
\graphicspath{{Fig/}{../Fig/}}
\usepackage{subcaption}
\usepackage{booktabs}
\usepackage{amssymb}
\usepackage{multicol}
\usepackage{amsmath}
\usepackage{amsthm}
\usepackage{lipsum}  
\usepackage{cite}
\usepackage{textcomp}
\def\BibTeX{{\rm B\kern-.05em{\sc i\kern-.025em b}\kern-.08em
    T\kern-.1667em\lower.7ex\hbox{E}\kern-.125emX}}

\usepackage{datetime}

\usepackage{color}

\providecommand{\comjh[1]}{\red{\com{JH}{#1}}}
\providecommand{\com[2]}{\begin{tt}[#1: #2]\end{tt}}

\newcommand{\bydef}{\stackrel{\Delta}{=}}

\newcommand{\beq}{\begin{equation}}
\newcommand{\eeq}{\end{equation}}
\newcommand{\beqa}{\begin{eqnarray}}
\newcommand{\eeqa}{\end{eqnarray}}
\newcommand{\beqan}{\begin{eqnarray*}}
\newcommand{\eeqan}{\end{eqnarray*}}
\newcommand{\bef}{\begin{figure}}
\newcommand{\enf}{\end{figure}}

\newcommand{\del}[1]{}

\definecolor{magenta}{cmyk}{0.1, 1, 0, 0}
\definecolor{greenedu}{cmyk}{1, 0, 1, 0.1}
\definecolor{cyanedu}{cmyk}{1, 0, 0, 0.1}
\newcommand{\edusug}[1]{{\color{magenta} #1}}

\newcommand{\bi}{\begin{itemize}}
\newcommand{\ei}{\end{itemize}}
\newcommand{\bc}{\begin{center}}
\newcommand{\ec}{\end{center}}
\newcommand{\ba}{\begin{array}}
\newcommand{\ea}{\end{array}}
\newcommand{\be}{\begin{equation}}
\newcommand{\ee}{\end{equation}}
\newcommand{\beno}{\begin{equation*}}
\newcommand{\eeno}{\end{equation*}}
\newcommand{\beqna}{\begin{eqnarray}}
\newcommand{\eeqna}{\end{eqnarray}}
\newcommand{\bd}{\begin{displaymath}}
\newcommand{\ed}{\end{displaymath}}
\newcommand{\beqnd}{\begin{eqnarray*}}
\newcommand{\eeqnd}{\end{eqnarray*}}

\renewcommand{\ni}{\noindent}

\newcommand{\cqfd}{\hfill \rule{2mm}{2mm}\smallbreak\indent}

\newtheorem{theorem}{\bf Theorem}[section]
\newtheorem{lemma}{\bf Lemma}[section]

\newtheorem{proposition}{\bf Proposition}[section]
\newtheorem{corollary}{\bf Corollary}[section]

\definecolor{red}{rgb}{1,0,0}

\definecolor{blu}{rgb}{0,0,1}

\definecolor{gre}{rgb}{0,0.7,0.3}

\begin{document}

\title{Necessary and sufficient conditions for the identifiability of isolated loops}

\author{Eduardo Mapurunga,
~~~Michel Gevers, \IEEEmembership{Life Fellow}
~~~Alexandre S.  Bazanella, \IEEEmembership{Senior Member}
\thanks{Eduardo Mapurunga is with the Data Driven Control Group, Department of Automation and Energy,
Universidade Federal do Rio Grande do Sul, Porto Alegre-RS, Brazil, eduardo.mapurunga@ufrgs.br.}
\thanks{Michel Gevers is with the Institute of Information and Communication Technologies, Electronics and Applied Mathematics (ICTEAM), UCLouvain, Louvain la Neuve, Belgium,
michel.gevers@uclouvain.be.}
\thanks{Alexandre S.  Bazanella is with the Data Driven Control Group, Department of Automation and Energy,
Universidade Federal do Rio Grande do Sul, Porto Alegre-RS, Brazil, bazanella@ufrgs.br.}
\thanks{This work  was supported in part by the Coordena{\c c}{\~ a}o de Aperfei{\c c}oamento de Pessoal de N{\'i}vel Superior - Brasil (CAPES) - Finance Code 001, by Conselho Nacional de Desenvolvimento Cient{\' i}fico e Tecnol{\' o}gico (CNPq),
by Wallonie-Bruxelles International (WBI), by a WBI.World excellence fellowship, and by a Concerted Research Action (ARC) of the French Community of Belgium.}
}

 \date{\empty}


 \maketitle


\begin{abstract}
				This Letter provides necessary and sufficient conditions on the excitation and measurement pattern (EMP) that guarantee identifiability of a dynamical network that has the structure of a loop. The conditions are extremely simple in their formulation, and they can be checked by visual inspection. They allow one to easily characterize all EMPs that make the loop network identifiable.
\end{abstract}


\section{Introduction}\label{intro}

Identification in dynamic networks has been subject of much research in the last ten years. Reference \cite{goncalves-warnick2-2008} is an early paper on this topic, inspired by the systems biology field. A network model and a framework for its study was later introduced by Paul Van den Hof and co-workers in the seminal paper \cite{van_den_hof_identification_2013}, which set the stage for the work on identification of dynamic networks. This Letter adopts the framework and network model from \cite{van_den_hof_identification_2013}.

What characterizes these dynamical networks  is that the signals are represented by nodes of the network, and these nodes are related to the other nodes through transfer functions. The identification task consists of identifying these transfer functions on the basis of the signals at the nodes and knowledge of the structure of the network, i.e. the topology of its corresponding graph. The identifiability questions could have been addressed on the  basis of the existing theory of closed loop identification.  However, this would have been a daunting task, given that, unlike classical closed loop systems,  networks are full of feedback loops  and parallel paths.  
The  paper \cite{van_den_hof_identification_2013} and the network model framework set up in that paper has opened up a whole new field, and not even ten years later a vast amount of challenging questions have been raised and many of them solved. Here is a very brief overview whose purpose is  to position the contribution of the present Letter in the sequence of developments.

In \cite{van_den_hof_identification_2013} it was assumed that all nodes were excited and measured. As a result, an input-output matrix of the network,  denoted $T(z)$, can be defined, which can always be identified from these data. The network identifiability question is then whether the network matrix, denoted $G(z)$  (whose elements are the transfer functions relating the nodes) can be recovered from this closed-loop transfer matrix $T(z)$. The paper \cite{van_den_hof_identification_2013} was followed by a flurry of contributions, in which the assumptions (all nodes are excited and measured) were progressively relaxed, and in which a range of new objectives were defined. Roughly speaking, these ranged from identification of the whole network (i.e. all elements of $G(z)$) to identification of some specific part of the network (a row of $G(z)$, for example), and to identification of a single element of $G(z)$
\cite{van_den_hof_identification_2013, gevers_identifiability_2017, dankers_identification_2016,  legat_local_2020, weerts_identifiability_2018, weerts_single_2018, everitt_empirical_2018, gevers_practical_2018}.
As for the assumptions on the signals, up to 2019, all contributions on the topic of identifiability of the network matrix $G(z)$ assumed that either all nodes were excited, or all nodes were measured.  A typical question would be: given that all nodes are excited, which nodes must be measured in order to identify the whole network? Or its dual version for the case where all nodes are measured.

Whatever the objective, the identifiability conditions were typically expressed as conditions on the rank of submatrices of the input-output transfer matrix $T(z)$. 
A major step forward was accomplished in \cite{bazanellaIdentifiabilityDynamicalNetworks2017, hendrickxIdentifiabilityDynamicalNetworks2019}, where the identifiability conditions were expressed in terms of properties of the associated graph, i.e. in terms of the topology of the network. This is a lot easier to check and to interpret than checking for the rank of transfer matrices. It required the introduction of the notion of {\it generic identifiability}, in order to cope with the possibility that a submatrix  might drop in rank on a thin subset of the parameter space. In \cite{hendrickxIdentifiabilityDynamicalNetworks2019} the assumption on the signals is that all nodes are excited but not all nodes are measured, the standard assumption (or its dual) until 2019, as stated above.

The first identifiability results for networks where not all nodes are excited AND not all nodes are measured were presented in \cite{Bazanella2019}. That paper first provided a necessary condition for identifiability of any network: each node must be either excited or measured, at least one node must be excited and at least one node measured.  The paper \cite{Bazanella2019} also  presented identifiability conditions for two special classes of networks, namely trees and loops. For a tree, it provided a necessary and sufficient condition for  identifiability. For the identifiability of an isolated loop, two sets of sufficient conditions were derived.  However, a necessary and sufficient condition remained elusive. Simple examples showed that these two sets of sufficient conditions were not necessary.

The specification of the combination of nodes that are excited and those that are measured defines the Excitation and Measurement Pattern (EMP), a term coined in \cite{mapurunga_optimal_2021} where the goal is to characterize all EMPs that yield identifiability of the considered network, or, even better a ``minimal EMP'' that yields identifiability. Here {\it minimal} means that one minimizes the sum of the number of excited and  measured nodes.
In \cite{mapurunga_identifiability_2021} minimal EMPs were characterized for some classes of acyclic networks with parallel paths, for which most nodes need to be excited and measured.

The contribution of the present paper is to derive necessary and sufficient conditions on the EMP for the identification  of an isolated loop. The practical importance of this result is twofold. The traditional way of looking at the identifiability of a network is to ask: is this network, with this specific pattern of excited and measured nodes identifiable? But a new paradigm, introduced in \cite{mapurunga_optimal_2021}, is to ask: what are all the EMPs that provide identifiability for this network? Our main theorem provides an easy answer to both questions for isolated loops.

In the last two years, several conjectures were proposed, including by the authors of this  Letter. These conjectures were typically expressed in rather complex terms. As it turns out, the necessary and sufficient conditions for the identifiability of an isolated loop, presented in Section~\ref{NSCloops}, are expressed in this Letter in very simple and explicit terms. They can be checked immediately by visual inspection of the distribution of excited and measured nodes along the loop. 

Essentially, our result states that an isolated loop is identifiable if and only if each node is either excited or measured, and in addition to that
				either there exists a node that is both excited and measured, 
				or
				the excited nodes (and hence also the measured nodes) are not all consecutive along the loop. 

%
%

The paper is organized as follows. In Section~\ref{loops} we pose the problem of identification of a dynamical network in general, and we then describe networks whose graph is a loop and give some of their key properties.
In Section~\ref{NSCconditions} we present our main result: necessary and sufficient conditions on the EMP for the identifiability of loop graphs. We also illustrate the ease of constructing valid EMPs as well as minimal EMPs. We conclude in Section \ref{conclusions}.

\section{Identification of isolated loops}\label{loops}

In this section we state the problem of the identification of an isolated loop within a dynamical network, and we derive a number of properties of an isolated loop in connection with this identification problem.

\subsection{The identification of a dynamical network}\label{subident}


The identification problem of a dynamical network consists of  identifying the elements of its network matrix $G(z)$, where the network is made up of $n$ nodes, with  node signals  denoted $\{w_1(t), \ldots, w_n(t)\}$,
and where  these node signals are related to each other and to  external excitation signals $r_j(t), j=1,\dots,m$ applied to a subset of its nodes by transfer functions. We denote by $\mathcal{W}$ the set of all $n$ nodes. The data available for this identification are the excitation signals, which are known, as well as the measures $y_{i}(t), i=1,\dots,p$ of a subset of the node signals.

Such network is described by the following
network equations, which we call the {\bf network model}:
 \beqna \label{model1}
 w(t) &=& G(z) w(t) + B r(t), \\
 y(t) &=& C w(t).
 \eeqna
The matrix $B$ is a binary selection matrix of size $n \times m$  having full column rank, and each of its columns contains one $1$ and $n-1$ zeros. The matrix $C$ is a binary selection matrix of size $p \times n$ having full row rank, and each of its rows contains one $1$ and $n-1$ zeros.
 These matrices define which of the $n$ nodes are excited and which are measured, respectively. We denote by   $\mathcal{B}$ the set of excited nodes, and by  $\mathcal C$ the set of measured nodes. This selection of excited and measured nodes is called the Excitation and Measurement Pattern (EMP).
The concept of EMP was introduced in \cite{mapurunga_optimal_2021}, where an EMP was called {\it valid} if it makes the network generically identifiable. It led to the concept of a {\it minimal EMP.}
A minimal EMP is one that guarantees generic identifiability of the network using the smallest possible number of excited and measured nodes \cite{mapurunga_optimal_2021}.

 To the network matrix one can associate a directed graph, in which a directed edge $(j,i)$ is present if $G_{ij}(z)\neq 0$. Thus, the graph defines the topology of the network, which is assumed to be known. The identifiability of the network matrix, of parts of a network matrix, or of a single edge within a network has been the object of  a large amount of publications within the last ten years, under a range of possible assumptions on the EMPs; the typical assumption until recently being that either all nodes are excited or that all nodes are measured \cite{hendrickxIdentifiabilityDynamicalNetworks2019, van_waarde_necessary_2019, cheng_allocation_2021}.

 In \cite{Bazanella2019}, the first results were presented for the identification of a network, or part of a  network,  with only partial excitation and measurement, i.e. where neither all nodes are excited nor all nodes are measured.
 The identifiability of the network matrix $G(z)$ for the case where a restricted set of nodes are excited and measured rests on the following relations.

From $G(z)$ one defines
 \be \label{Tdef}
 T(z)\bydef (I - G(z))^{-1}.
 \ee
The {\bf input-output model} corresponding to the network model (\ref{model1})   is then given by
 \be \label{Mdef}
 y(t)=M(z)r(t) ~~~\mbox{with}~~~M(z) \bydef  CT(z)B.
  \ee

To keep things simple, we  assume  that the vector $r(t)$ is sufficiently rich so that, for all choices of $C$ and $B$, $M(z)$ can be consistently estimated by standard open loop MIMO (Multiple Input Multiple Output) identification techniques from $\{y(t), r(t)\}$ data.
The question of identifiability of the network matrix $G(z)$ from the given data is then equivalent to the question of whether $G(z)$ can be generically recovered from the known $M(z)$  (i.e. from $CT(z)B$), knowing that $G(z)$ is related to $M(z)$ by (\ref{Tdef})-(\ref{Mdef}).

Generic identifiability of a network matrix was defined in \cite{hendrickxIdentifiabilityDynamicalNetworks2019}. Simply stated, the network matrix $G(z)$ is called {\it generically identifiable} with a given EMP if, for any rational parametrization $G(P,z)$ consistent with its associated graph, $G(P,z)$ can be uniquely recovered from $M(z)$ except possibly on a zero measure set in the space of parameters $P$. See \cite{hendrickxIdentifiabilityDynamicalNetworks2019} for details and an example.


 Among other results, \cite{Bazanella2019} provided the following necessary condition for the identification of the network matrix when only partial excitation and measurement is available.

\begin{proposition}\label{neccond1}
 The network matrix $G(z)$ is generically  identifiable from excitation signals applied to $\cal B$ and measurements made at $\cal C$ only if $\cal B \neq  \emptyset$,  $\cal C \neq  \emptyset$ and
 $\cal B \cup \cal C = \cal W$. \end{proposition}
This result provides necessary conditions on the valid EMPs: each node must be either measured or excited,
 which implies that the smallest number of excitations and measurements that a minimal EMP could have is $n$.

 \subsection{Loop graphs and their properties}\label{loopgraphs}

  A ``loop graph" is a graph that consists of a single loop and nothing more. Its network matrix is in the form
\begin{equation}\label{laco}
G =
\left[\begin{array}{ccccc}
	0 & 0 & \ldots & 0 & G_{1n} \\
	G_{21} & 0 & \ldots & 0 & 0 \\
	0 & G_{32} & \ldots & 0 & 0 \\
	\vdots & & \edusug{\ddots} & & \vdots \\
	0 & 0 & \ldots & G_{n(n-1)} & 0 \\
\end{array}\right].
\end{equation}

The loop we are interested in can be a graph by itself, as in (\ref{laco}), or part of a larger graph. When the loop of interest is part of a larger graph, some of its nodes may belong to other loops. 
When this is not the case - that is, no other loop in the graph
contains any of the nodes of the loop of interest - we will say that it is an {\em isolated loop}. All results in this paper pertain to the identifiability of isolated loops.

The necessary condition ${\cal B} \cup {\cal C }={ \cal W}$ of Proposition~\ref{neccond1}, which applies to all dynamic networks,  means that all nodes must be involved in the identification
process: they must be either measured or  excited. When it comes to loops, the question we address is whether any EMP that satisfies that necessary condition would make $G(z)$ identifiable  (it would then be a  minimal  EMP with smallest cardinality\footnote{Cardinality of an EMP is defined as: $ \mid{\cal B}\mid +  \mid{\cal C}\mid  $.}), or whether additional conditions apply.

Our main result about identifiability of loops in \cite{Bazanella2019} showed that adding a single
measurement or a single excitation to such  EMP is sufficient to achieve generic identifiability of an isolated loop.

\begin{proposition}\label{sufloop}
All transfer functions in an isolated loop  
are generically identifiable if  ${\cal B} \cup {\cal C} = {\cal W}$ and ${\cal{B} \cap \cal{C} }\neq \emptyset$.
\end{proposition}

In order to prove the main result of this paper, namely necessary and sufficient conditions for the identifiability of an isolated loop, we now derive some properties of loops. We first recall some expressions and properties derived in the proof of Theorem V.2 of \cite{Bazanella2019}.

Assume, without loss of generality, that the node indices in the loop go from $1$ to $n$ and that
the arrows go from $i$ to $i+1$ in the cycle, as in (\ref{laco}).
Define the product of all transfer functions in the loop as follows:
\be \label{Pdef}
P \bydef G_{1n}G_{n,n-1}\ldots G_{32}G_{21}.
\ee

Observe that the closed loop transfer function from one node to itself is
\begin{equation}\label{eq:T=R=1-P}
T_{ii}=R := (1-P)^{-1}.  
\end{equation}
For distinct $i,k$, we also define
\begin{eqnarray}
 P_{ik} \bydef G_{i,i-1}G_{i-1,i-2}\dots G_{k+1,k} ~~ \text{if} ~~k<i, && \label{Pik1} \\
 P_{ik}  \bydef G_{i,i-1}G_{i-1,i-2}\ldots G_{1n}G_{n,n-1}\ldots  G_{k+1,k} ~ \text{if} ~ k>i . && \label{Pik2}
\end{eqnarray}

The next Lemma provides relations between the quantities $P, P_{ik}, T_{ik}$ and $ G_{ik}$ that will be useful for proving our main result.

\begin{lemma}\label{relPTG}
The transfer functions $P, P_{ik}, T_{ik}$ and $ G_{ik}$ are related by the following expressions for any $i, k, j$:
\begin{eqnarray}
T_{ik} &= &P_{ik}R,  \label{eq:T=PR}\\
P&=&  P_{ki}P_{ik},  \label{eq:P=prodP}\\
G_{i+1,i} &=& \frac{P_{ji}}{P_{{j,i+1}}} = \frac{P_{i+1, j}}{P_{i j}} = \frac{T_{ji}}{T_{{j,i+1}}} = \frac{T_{i+1,j}}{T_{{ij}}}. \label{eq:GPT}
\end{eqnarray}
\end{lemma}
\ni Proof: Expressions (\ref{eq:T=PR}) and (\ref{eq:P=prodP}) follow immediately from the input-output relationship and definition of $P_{ik}$.
				The expressions in  (\ref{eq:GPT}) can be verified by direct computation.
				For $j > i + 1$, we have
				\beqnd
								\frac{P_{ji}}{P_{j, i + 1}} = \frac{ G_{j, j-1} G_{j-1, j-2} \cdots G_{i+2, i+1} G_{i+1, i}}{ G_{j, j-1} G_{j-1, j-2} \cdots G_{i+2, i+1}} = G_{i+1, i} 
				.\eeqnd
				When $i > j$, the same relationship can be obtained using (\ref{eq:P=prodP}).
				\cqfd

With these expressions under our belt, we are now ready to prove our main result.

\section{Necessary and sufficient conditions for identification of  loops}\label{NSCloops}

We first consider the special case where the loop has either 2 nodes or 3 nodes, i.e. $n=2$ or $n=3$.

\begin{theorem}\label{2or3loops}
All transfer functions  in an isolated
loop with $n\leq 3$ are generically identifiable if and only if  ${\cal B} \cup {\cal C} = {\cal W}$ and ${\cal{B} \cap \cal{C} }\neq \emptyset$.
\end{theorem}
\ni Proof:

For $n=2$ we have:
				\begin{align}
								T = \left[\begin{matrix}
								\frac{1}{1 - G_{12} G_{21}} &  \frac{G_{12}}{1- G_{12} G_{21} }\\
								\frac{G_{21}}{1- G_{12} G_{21}} & \frac{1}{1- G_{12} G_{21}}
				\end{matrix}\right]
				\label{eq:T_2}
				.\end{align}
				For $n = 3$ we have:
				\begin{align}
								T = \left[\begin{matrix}
								\frac{1}{1- G_{13} G_{21} G_{32}} & \frac{G_{13} G_{32}}{1 - G_{13} G_{21} G_{32}} & \frac{G_{13}}{1 - G_{13} G_{21} G_{32} }\\
								\frac{G_{21}}{1 - G_{13} G_{21} G_{32}} & \frac{1}{1 - G_{13} G_{21} G_{32}} & \frac{G_{13} G_{21}}{1 - G_{13} G_{21} G_{32}}\\
								\frac{G_{21} G_{32}}{1 - G_{13} G_{21} G_{32} } & \frac{G_{32}}{1 - G_{13} G_{21} G_{32}} & \frac{1}{1 - G_{13} G_{21} G_{32}}
								\end{matrix}\right]
								\label{eq:T3}
				.\end{align}

Sufficiency follows directly from Proposition~\ref{sufloop} above, which has been proven in \cite{Bazanella2019}. For necessity, inspection of the $T$ matrix shows that  if no node is excited and measured, then for $n=2$, $T$ contains only a single known element (impossible to recover 2 $G_{ij}$). For $n=3$, there are only 2 independent elements of $T$ (impossible to recover 3 $G_{ij}$).
\cqfd

What this Theorem shows is that for an EMP to guarantee generic identifiability of a loop with less than four nodes it must have at least one node excited and measured, while the others must be either excited or measured.
A minimal EMP is therefore characterized by one node being excited and measured, which results in a total of four minimal EMPs for $n = 2$ and twelve minimal EMPs for $n = 3$.

We now consider isolated loops that have at least 4 nodes.
In \cite{Bazanella2019}, a sufficient condition for identifiability had been derived for loops that have an even number of nodes, larger than 3. It was shown that identifiability is achieved if the EMP obeys the following interleaving condition between excited and measured nodes.

\begin{proposition}\label{theorem_even}
Let $n$ be even and larger than 3. All transfer functions in an isolated  loop can be identified if its nodes are alternately measured and excited.
\end{proposition}

This interleaving condition on the  EMP is clearly not necessary for identifiability of an isolated loop, as follows from Proposition~\ref{sufloop}, but it has inspired the development of the main theorem of this paper, which provides necessary and sufficient conditions for identifiability.

\begin{theorem}\label{NSCconditions}
All transfer functions in an isolated loop are generically
identifiable if and only if ${\cal B} \cup {\cal C} = {\cal W}$ and, in addition:
(i) either ${\cal{B} \cap \cal{C} }\neq \emptyset$,
 or
(ii) there exist at least two measured nodes in the loop, each of which is immediately followed by an excited node.
\end{theorem}
\ni {\bf Proof.}
That each node must be excited or measured follows from Proposition~\ref{neccond1}.\\
\ni {\it Proof of sufficiency.}\\
If (i) holds, the loop is identifiable by Theorem V.2 of \cite{Bazanella2019}.
Consider now that (ii) holds. Without loss of generality let nodes $k$ and $n$ be the two measured nodes that are followed immediately by nodes $k+1$ and $1$, which are excited. From these measurements and excitations we obtain: $T_{k 1}, T_{n 1}, T_{n, k+1}$, and $T_{k, k+1}$, and we can form the product:
$$\frac{T_{n 1} T_{k, k+1}}{T_{k 1} T_{n, k+1}} =
					\frac{P_{n 1} P_{k, k+1}}{P_{k 1} P_{n, k+1}} = P_{n k} P_{k n} = P, $$
where the equalities follow from Lemma~\ref{relPTG}. Once $P$ is known, the transfer functions $G_{{ij}}$ can be calculated step by step from the available $T_{ij}$, remembering that each node in the loop is either measured or excited.

Suppose node $2$ is excited, we can recover $G_{21} = T_{k1}/T_{k2}$.
If $2$ is measured, we can identify  $G_{21} = T_{21}(1 - P)$.
Consider that the next node $3$ is excited, then we can recover $G_{32} = T_{k1}/(G_{21}T_{k3})$.
If $3$ is measured, we identify $G_{32} = T_{31}(1 - P)/G_{21}$.
Proceeding in a similar fashion we can recover $G_{i, i-1}$ from knowledge of the previous identified modules as:
\begin{align}
			G_{i, i-1} = \frac{T_{n1}}{P_{i-2, 1} T_{n i}}, ~ \text{if} ~ i \in {\cal B}, \\
			G_{i, i-1} = \frac{T_{i1}(1 - P)}{P_{i-1, 1}}, ~ \text{if} ~ i \in {\cal C},
\end{align}
for $i = 3, \dots, n$. The last transfer function can be recovered from $P$ since $G_{1n} = P / P_{n 1}$.

\ni {\it Proof of necessity.}\\
Consider that no node is both excited and measured. We show that condition (ii) must then hold to guarantee identifiability.
If only one node is excited, then we can identify only $n-1$ transfer functions $T_{ij}$; hence we cannot identify the $n$ elements  $G_{ij}$.
The same holds if only one node is measured.
This shows that we need at least two excited and two measured nodes in the loop.
Suppose now that the loop contains at least two measured nodes and two excited nodes, but that condition (ii) does not hold. This implies that all the measured nodes are consecutive, and so are all the excited nodes. Without loss of generality, let nodes $1$ to $k$ be the excited nodes and let $k+1$ to $n$ be the measured nodes.

Then, with $P_{ik}$ defined in (\ref{Pik1}) for $i>k$,  the corresponding matrix $M(z)\bydef CT(z)B$ defined in (\ref{Mdef})  has the following form.
\beqnd
&&M(z) \bydef CT(z)B\\
&&=R\left(\begin{array}{ccccc}
P_{k+1,1} & P_{k+1,2} & \ldots & P_{k+1,k-1} & P_{k+1,k} \\
P_{k+2,1} & P_{k+2,2}& \ldots & P_{k+2,k-1} & P_{k+2,k} \\
\vdots & \vdots&   & \vdots & \vdots \\
P_{n1} & P_{n,2} & \ldots & P_{n,k-1} & P_{nk}\end{array}\right)
.\eeqnd
It now follows from (\ref{eq:GPT}) that the first row of $M(z)$ allows one to successively compute $G_{21}, G_{32}, \ldots, G_{k,k-1}.$ It follows from (\ref{eq:GPT}) that all elements of the second row of $M(z)$ are equal to the corresponding elements of the first row multiplied by $G_{k+2,k+1}$. Thus, knowledge of the second row allows one to compute one additional element of $G(z)$, namely $G_{k+2,k+1}$. Pursuing row by row downwards up  to the last row shows that we can compute $G_{k+2,k+1}, \ldots,G_{n,n-1} $. Collecting these results shows that, with this EMP (i.e. the loop consists of $k$ consecutive excited nodes followed by $n-k$ measured nodes), the corresponding
$CT(z)B$ allows one to identify the edges $G_{21}, G_{32},\dots G_{k,k-1}$ as well as the edges $G_{k+2, k+1}, \dots , G_{n,n-1}$  can be identified, but not the edges   $G_{k+1, k}$  and $G_{1 n}$.
\cqfd

We assume from now on that the loops we consider have at least $3$ nodes, noting that a loop with two nodes is just a simple feedback system for which the identifiability conditions are well established. The following Corollary provides an alternative formulation for the results of Theorem~\ref{2or3loops} and 
Theorem~\ref{NSCconditions} which yields an even simpler way of checking the identifiability of an isolated loop.

 \begin{corollary}\label{NSCcondCor}
All transfer functions in an isolated loop are generically
identifiable if and only if ${\cal B} \cup {\cal C} = {\cal W}$ and, in addition:
(i) either ${\cal{B} \cap \cal{C} }\neq \emptyset$,
 or
(ii) the excited nodes (and hence also the measured nodes) are not all consecutive along the loop. 
\end{corollary}
\ni {\bf Proof: } the result is included in the proof of Theorem~\ref{NSCconditions}. For the special case of $n=3$, it is easy to see that if ${\cal B} \cup {\cal C} = {\cal W}$ and ${\cal{B} \cap \cal{C} }= \emptyset$, then necessarily condition (ii) is violated.
\cqfd

Not only are the conditions 
of  Corollary~\ref{NSCcondCor} necessary and sufficient, but in addition  their verification on a given loop graph is trivial: it can be done by visual inspection. Conversely, if the objective is to establish minimal EMPs for the identification of a loop graph, condition (ii) in this Corollary  provides all minimal EMPs.
There is a total of $2^n - n(n-1) - 2$ minimal EMPs from which the user can choose from.
			Once the minimal EMPs are characterized, all other valid EMPs can be obtained by just picking at least one node to be both excited and measured.
			These EMPs correspond to a number of $\sum_{k = 1}^{n - 1} 2^{n - k} n! /(k!(n - k)!) + 1$ out of the total EMPs. 

\begin{figure}[h!]
							\centering
							\caption{Two possible EMPs for a 5-node loop. {\bf E} stands for an excited node and {\bf M} for a measured node.}
							\label{fig:exampleloop5nodes}
							\begin{subfigure}{0.5\columnwidth}
											\centering
							     \includegraphics[width=0.8\columnwidth]{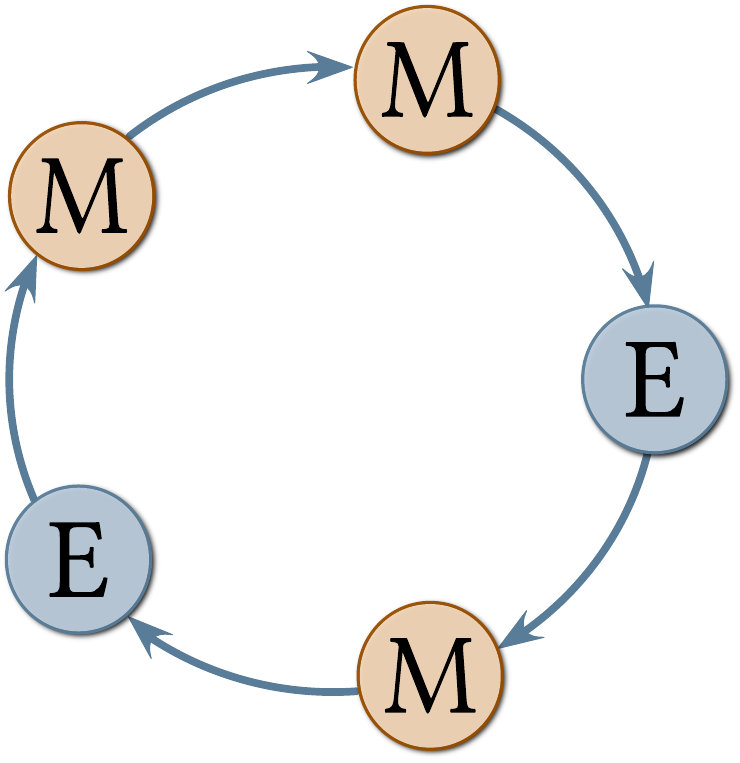}
									 \caption{A minimal EMP. }
							\end{subfigure}~
							\begin{subfigure}{0.5\columnwidth}
											\centering
							     \includegraphics[width=0.8\columnwidth]{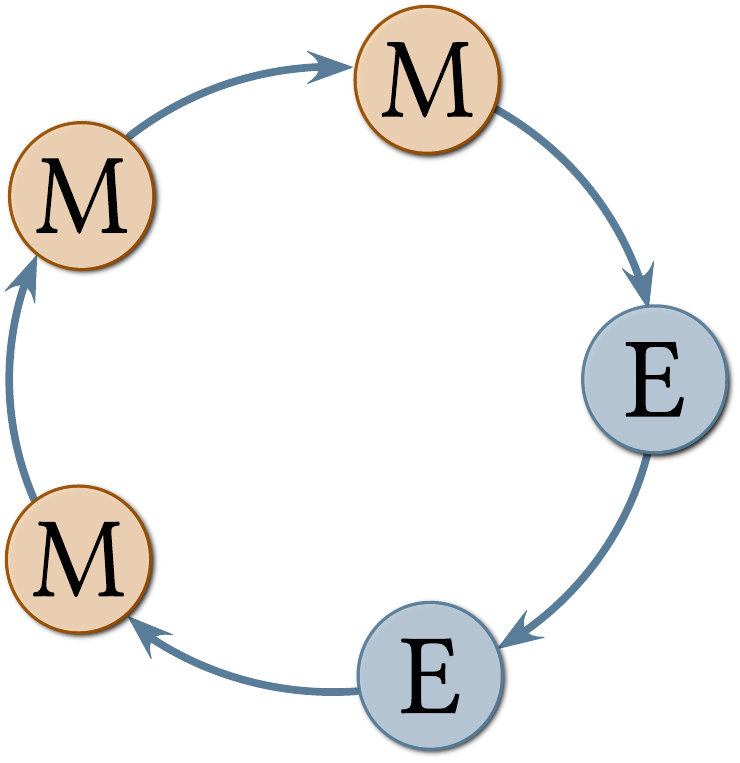}
									 \caption{Not valid EMP.}
									 \label{subfig:niEMP}
							\end{subfigure}
			\end{figure}

It is easy to spot invalid EMPs. Consider, for instance, the two EMPs for a 5-node loop depicted in Figure \ref{fig:exampleloop5nodes}, where the one on the left is minimal and the one on the right is not valid.
An EMP satisfying the necessary conditions of
Proposition \ref{neccond1}
is {\em invalid} only if  no node in the loop is both excited and measured,
and  all excitations and measurements are arranged in an uninterrupted sequence
as in Figure \ref{subfig:niEMP};  in other words, the excited nodes (and therefore
the measured nodes) are all contiguous.

%

When we consider all possible EMPs, the number of valid EMPs outnumber those that are not generically identifiable for any number of nodes in the loop, since it is sufficient to have at least one node excited and measured with any combination of the remaining nodes.
This means that if the user were to choose randomly an EMP, it is likely that the chosen EMP is identifiable. Table I is a clear illustration of this. It provides, for loops having 2 to 10 nodes, the number of minimal EMPs, of valid EMPs, and of invalid ones, assuming that in all cases the necessary condition is satisfied, i.e. each node is either excited or measured.


\begin{table}[h!]
				\centering
				\caption{Number of minimal/valid/invalid EMPs for a loop where each node is at least excited or measured.}
				\label{tab:ratio}
				\begin{tabular}{c c c c}
								\toprule
								nodes & minimal EMPs & valid EMPs & invalid EMPs \\
								\midrule
								2 & 4 & 5 & 2 \\
								3 & 12 & 19 & 8 \\
								4 & 2 &  67 & 14 \\
								5 & 10  & 221 & 22 \\
								6 & 32  & 697 & 32 \\
								7 & 84  & 2143 & 44 \\
								8 & 198  & 6503 & 58 \\
								9 & 438  & 19609 & 74 \\
								10 & 932  & 58957 & 92 \\
							\bottomrule
				\end{tabular}
\end{table}


\section{Conclusions}\label{conclusions}

We have derived necessary and sufficient conditions for the generic identifiability of a loop network. The conditions are very simple to check by visual inspection of the corresponding graph. Unlike the other  results on identifiability of networks so far, no recourse to rank conditions or to vertex disjoint paths are required. Besides the necessary requirement that each node must be either excited or measured, as is the case for all network structures, the requirement for loops with 4 nodes or more is that either one node is both excited and measured, or that  not all measured (and hence not all excited) nodes are contiguous.
Our results make it extremely easy to choose an EMP that makes the network identifiable and that is convenient for the user from a practical point of view.

\bibliographystyle{IEEEtran}
\bibliography{StrucDyNetB}


\end{document}